\documentstyle[12pt]{article}
\date{}
\newcommand{\beq}{\begin{equation}}
\newcommand{\eeq}{\end{equation}}
\newcommand{\beqn}{\begin{eqnarray}}
\newcommand{\eeqn}{\end{eqnarray}}
\newcommand{\dst}{\displaystyle}
\newcommand{\fr}[2]{\frac{{\dst #1}}{{\dst #2}}}
\newcommand{\epe}{\mbox{$e^+e^-\,$}}
\newcommand{\ggam}{\mbox{$\gamma\gamma\,$}}

\begin{document}

\title
{ The $e^+e^-$ pair production at $\mu^+\mu^-$ collider} 
\author{Ilya F.Ginzburg\\Institute of Mathematics.
  630090. Novosibirsk. Russia\\ E-mail: ginzburg@math.nsk.su}
\maketitle
\begin{abstract}
The main source of \epe pair at $\mu^+\mu^-$ collider is
incoherent process $\mu^+\mu^-\to \mu^+\mu^-\epe$ with cross
section about 10 mb (at muon energy 2 TeV). The corresponding
distributions are discussed briefly. These pairs give the
dangerous background. 
 
The coherent production of \epe pairs here is negligibly small. 
\end{abstract}

The possible $\mu^+\mu^-$ collider is discussed widely now (see
e.g. \cite{Palm}) with the expected set of parameters:
$$
E=2 \mbox{ TeV }(\gamma_{\mu}=2\cdot 10^4);\quad L=2\cdot
10^{35}\mbox{ cm}^{-2}s^{-1};
$$
\beq
N=2\cdot 10^{12}; \sigma_{x,y}=3\mu m;\;\sigma_z=3 mm;\; 
f_{rep}=30s^{-1};\;B\sim\fr{eN}{2\sigma_x\sigma_z}\sim 0.5\cdot 
10^3\;T.\label{param}
\eeq
Here $B$ is the characteristic magnetic field of bunch at the
collision point. 

The \epe pair production in this collision seems to be source of
dangerous background \cite{Pisin}. We consider two mechanisms of
this production --- standard {\em incoherent production},
$\mu^+\mu^-\to\mu^+\mu^-\epe$, and {\em coherent production} on
the collective electromagnetic field of the oncoming bunch.

\section{Incoherent production, $\mu^+\mu^-\to \mu^+\mu^-\epe$}

The production of \epe pairs at the collision of charged
particles was considered first 60 years ago \cite{LL}. The
asymptotical behaviour of total cross section was obtained in
this paper. The numerous papers related to the more detail and
precise treatment of this process are reviewed in ref.
\cite{BGMS}.

The similar process $pp\to pp\epe$ have been considered in
details in ref. \cite{lum}. Both the precise equations and the
detailed qualitative description of the main features of process
with numerical estimates were presented in this paper (to use
this process for the luminosity monitoring at ISR).  The
equations there are valid for our case with the evident
renotations, some numerical estimates should be reconsidered due
to much higher Lorentz--factor in our case. 

We present here some general characteristics of the process for
the preliminary rough estimates.

1) The total cross section of the process is \cite{Rac}:
\beq
\sigma=\fr{28\alpha^4}{27\pi m_e^2}\left[(l-2.12)^3 +2.2(l-2.12)
+0.4 \right]\approx 10\; mb .\label{total}
\eeq

2) The main mechanism of process is two--photon production of
\epe pairs via collision of two virtual photons, emitted by
muons. Main features of process can be obtained with the
equivalent photon (Weizsacker--Williams) approximation (EPA).
The spectra of virtual photons are obtained from the analysis of
Feynman diagrams. Their dependence on the energy $\omega$ and
virtuality $Q^2$ has the form
\beqn
&dn(\omega ,Q^2)&=\fr{\alpha}{\pi}\fr{dx}{x}\fr{dQ^2}{Q^2}\left
[1-x+\fr{x^2}{2}- (1-x)\fr{Q^2_{min}}{Q^2}\right]; \quad
x=\fr{\omega}{E};\label{spectr}\\
&& Q^2\geq Q^2_{min}=m_{\mu}^2\fr{x^2}{1-x}.\label{min}
\eeqn
Last inequality is obtained easily from kinematics.

The cross section of subprocess $\ggam\to \epe$ has maximum at
the effective mass squared of produced system $W^2\approx
8m_e^2$. Therefore, the effective mass of produced $e^+e^-$
system is near the threshold, the transverse momenta of produced
particles are $\sim m_e$. Besides, this cross section decreases
quickly with the growth of virtuality above $m_e^2$. In other words,
the main contribution to the cross section is given by the region
$$
m_e^2>Q^2>Q^2_{min}.
$$
Using eq. (\ref{min}), we obtain from here limitation for the
energy of photon for the process:
\beq
x=\fr{\omega}{E}< \fr{m_e}{m_{\mu}}\Rightarrow \omega < \gamma
m_e. \label{limen}
\eeq
Therefore, --- in accordance with the "naive" expectations ---
the Lorentz--factor of produced \epe pair cannot be higher than
that of the initial muon.

Besides, the number of equivalent photons for this production is
obtained by integration of eq. (\ref{spectr}) over virtuality, it
is
\beq
dn(\omega)=\fr{2\alpha}{\pi}\fr{dx}{x}\left[\ln
\left(\fr{m_e}{m_{\mu}x}\right) -\fr{1}{2}\right] .\label{spectren}
\eeq
Note, that this quantity is much lower than that for the
two--photon production of muons or hadrons, which is
\beq
\sim(2\alpha/\pi)(dx/x)\ln(1/x).\label{wrsp}
\eeq
The source of this difference is the much higher upper limit of
effective virtualities for these processes, which determined by
the much higher scale of the $Q^2$ dependence for these
subprocess. (For more detail discussion see ref. \cite{BGMS}.)

3) The produced pairs distributed uniformly in the rapidity
scale. The distribution over the total energy of pair $\epsilon$ is
\beq
d\sigma=\left\{\begin{array}{c c l}
\fr{56\alpha^4}{9\pi m_e^2} \fr {dk_z}{\epsilon}
\left[\ln^2\gamma - \ln^2\fr{\epsilon}{m}\right] &\mbox { at }&
\epsilon<\gamma m_e;\\
\sim\fr{\alpha^4}{m_e^2}\fr {d\epsilon}{\epsilon}
\left(\fr{\gamma m_e}{\epsilon}\right)^2 &\mbox { at }&
\epsilon>\gamma m_e .
\end{array}\right .
\eeq
Here $k_z$ is the longitudinal momentum of the pair, $|k_z|\approx
\epsilon$.

Mean energy of pair is $\sim 2m_e\gamma/\ln{\gamma}\sim 2$
GeV. In accordance with eq. (\ref{total}), the number of pair
produced is about $ 10^5$ per bunch collision, i.e. about $10^8$
during the life of bunch. Therefore, the entire energy losses
due to the discussed process are about $2\cdot 10^{-6}\%$, i.e.
negligible.

4) The distribution over the energy of one electron $\epsilon_1$,
emitted along the motion of initial $\mu^+$, has the form
\beq
d\sigma=\fr{56\alpha^4}{9\pi m_e^2} \fr {dk_{1z}}{\epsilon_1}
\left[\ln^2\gamma - \ln^2\fr{\epsilon_1}{m}\right]\quad  (\epsilon_1\gg
m_e). \label{endistr}
\eeq
Here $k_{1z}$ is the longitudinal momentum of the electron. 
The effective mass of produced pairs is near the threshold and
their total transverse momentum is very low. Therefore, the main
part of produced electrons moves initially precisely along the
beam. 

However, as it was pointed out by Palmer \cite{PPalm}, the
created electrons are deflected by the magnetic field of opposite
beam. It is easily seen that the electrons with the energy
$<100$ MeV have Larmor radius $R=\epsilon_1/B$ less than 1 mm,
and they are invisible in the detectors. The electrons with the
energy $100$ MeV $<\epsilon_1 < 1$ GeV have Larmor radius between
1mm and 1cm, they are dangerous for the vertex detectors.  Some
part of these electrons can reach main detector. In accordance
with eq.  (\ref{endistr}), the corresponding cross section is
about 1 mb for each direction (along $\mu^-$ and along $\mu^+$),
we expect about 10 000 of these electrons (or positrons) per
bunch crossing. Last, electrons with highest energy $1$ GeV
$<\epsilon_1 < 10$ GeV are deflected at the angle $\beta=
e^2N/(2\sigma_x\epsilon_1)>50$ mrad, to the main detector. The
corresponding cross section is about 0.4 mb for each direction,
with 4 000 electrons and total energy flux about 10 TeV per bunch
crossing.
 
The more detail studies of these problems are necessary.  

\section{Coherent production}

The coherent production was considered in the paper \cite{Pisin}.
It is based on the following facts: The electromagnetic field of
bunch $B$ is rather large. Its ratio to the Schwinger critical
field for the creation of \epe pairs from vacua $B_c$ in our case
is small, $B/B_c\approx 2\cdot 10^{-7}$. But the quantity 
$(E/m_e)(B/B_c)\sim 0.4$.

The idea of the paper \cite{Pisin} looks very nice: Photons
(either real or virtual) from one bunch which traverse the
electromagnetic field of other bunch would turn into \epe pairs.
The rate of this process is determined by the parameter (related
to electrons only!)
\beq 
\chi=\fr{\omega}{m_e}\fr{B}{B_c}\quad \left(B_c=\fr{m_e^2
c^3}{e\hbar} \approx 4.4\cdot 10^{13}\; G\right).\label{chi}
\eeq
The probability of pair creation per unit time can be written as 
($\hbar=c=1$) \cite{Ritus}
\beq
\fr{3\sqrt{3}}{16\sqrt{2}}\fr{\alpha m^2}{\omega}\chi \exp\left(
-\fr{8}{3\chi}\right).\label{prob}
\eeq

The photons with energy $\omega\approx E$ with $\chi\sim 0.4$
exist in the spectrum of equivalent (virtual) photons, for these
photons large enough production of \epe pairs is expected. The
number of produced \epe pairs is obtained by convolution of this
probability with the spectrum (\ref{wrsp}).

Unfortunately, the last point is inexact. Indeed, the spectrum
of equivalent photons depends strongly on the nature of both
particle created and system produced. The simple equation
(\ref{spectr} for the spectrum is valid at $Q^2<Q^2_{max}$. The
quantity $Q^2_{max}$ determined in our case by the nature of
subprocess considered, it is $ \hat s/4$ where $\hat s$ is the
effective mass of the produced system. The main contribution is
from region near the threshold. Therefore, $Q^2_{max}\sim
m_e^2$. (High value of $Q^2$ corresponds to small size of
production region.)

At $Q^2>Q^2_{max}$ the quantity $dn$ decreases quickly, the
contribution of this region into a flux of equivalent photos is
negligible. The inequality $Q^2_{max}<Q^2_{min}$ results in
limitation for the virtual photon energy (\ref{limen}) and
spectrum (\ref{spectren}), which differs strongly from the
spectrum (\ref{wrsp}). For this spectrum the parameter $\chi$ is
limited from above by the value $\chi_{max}=\gamma{B\over B_c}
\approx 0.002.$ It means that the coherent production of
\epe pairs is negligibly small.

The same very conclusion can be explained by other words:

The spectrum (\ref{wrsp}) has no relation to the \epe
production. In fact, the virtual photon is described by both
energy $\omega$ and virtuality $Q^2\geq Q^2_{min}$
(\ref{spectr},\ref{min}). The virtuality $Q^2$ (if it is higher
than $m_e^2$) determine the scale of production region, so that
for such photon the critical field for \epe pair production
become $\sim(Q^q/m_e^2)B_c$. The more precise interpolation
equation, obtained from the analysis of corresponding dependence
for the \epe production on real photon, is
\beq
B_c\Rightarrow \fr{m_e^2+Q^2/6}{m_e^2} B_c. \label{crit}
\eeq
The probability of \epe coherent production should be obtained by
the convolution of the spectrum (\ref{spectr}) with the
probability (\ref{prob}) related to this critical field.  Using
for $Q^2$ its least value for given photon energy (\ref{min}),
we obtain that the maximal value of the quantity $\chi$ is
reached at $x\approx 2.5(m_e/m_{\mu})$, it is $\sim\gamma
(B/B_c)\approx 0.002$.  

\vspace{0.3cm}

I am very thankful to K.J.~Kim, N.~Mokhov, R.~Palmer, A.~Sessler,
A.~Skrinsky V.~Serbo, K.~Yokoya, M.~Zolotarev for useful
discussions. This work is supported by grants of Russian
Foundation of Fundamental Investigations and INTAS--93--1180.

\end{document}